\documentclass[fleqn,usenatbib]{mnras}

\usepackage{newtxtext,newtxmath}
\usepackage[T1]{fontenc}
\usepackage{ae,aecompl}

\usepackage{graphicx}	% Including figure files
\usepackage{amsmath}	% Advanced maths commands
\usepackage{amssymb}	% Extra maths symbols

\title[HRS + HCI with ELT : looking for O$_2$ in Proxima b]{High Resolution Spectroscopy and High Contrast Imaging with the ELT : looking for O$_2$ in Proxima b}

\author[G. A. Hawker et al.]{
George A. Hawker,$^{1}$\thanks{E-mail: gah43@cam.ac.uk (GAH)}
Ian R. Parry,$^{1}$
\\
$^{1}$Institute of Astronomy, University of Cambridge, Madingley Road, Cambridge, UK, CB3 0HA\\
}

\date{Accepted 2019 January 27. Received 2019 January 22; in original form 2018 August 14}

\pubyear{2015}
\hypersetup{draft}
\begin{document}
\label{firstpage}
\pagerange{\pageref{firstpage}--\pageref{lastpage}}
\maketitle

% Abstract of the paper
\begin{abstract}
We research the requirements of High Contrast Imaging when combined with the cross correlation (CC) of high resolution spectra with known spectroscopic templates for detecting and characterising exoplanets  in reflected light. We simulate applying the technique to a potentially habitable  Proxima b-like planet and show that the O$ _2$ A-band spectral feature could feasibly be detected on nearby rocky exoplanets using future instruments on the ELT . The technique is then more widely analysed showing that detections of planets and O$ _2$ with signal to noise in the CC function ( SNR$_{CC}$) $ >3 $ can be obtained when the signal to noise of the simulated  planet spectrum ( SNR$_{spec}$) is from 0.25 to 1.2. We place constraints on the spectral resolution, instrument contrast, point spread function (PSF), exposure times and systematic error in stellar light subtraction for making such detections.  We find that accurate stellar light subtraction (with 99.99\% removal) and PSFs with high spatial resolutions are key to making detections. Lastly a further investigation suggests the ELT could potentially discover and characterise planets of all sizes around different spectral type stars, as well as detecting O$ _2 $ on Super-Earths with habitable zone orbits around nearby M stars.
\end{abstract}

% Select between one and six entries from the list of approved keywords.
% Don't make up new ones.

%%%%%%%%%%%%%%%%%%%%%%%%%%%%%%%%%%%%%%%%%%%%%%%%%%

%%%%%%%%%%%%%%%%% BODY OF PAPER %%%%%%%%%%%%%%%%%%

\section{Introduction}
The ability to both detect and characterise extrasolar planets by their reflected light is currently one of the most sought after goals in modern astrophysics. Current methods of detection are heavily restricted to either young, self-luminous, giant planets (Direct Imaging) or closely orbiting, short orbital period planets (transit and radial velocity methods), thus potential methods to expand the observable parameter space of the exoplanet population need thorough investigation. We investigate one such method which combines High  Resolution Spectroscopy (HRS) with High Contrast Imaging (HCI) building upon the idea proposed by \cite{2002ApJ...578..543S} and further work of \cite{2015A&A...576A..59S}, \cite{wang2017} and \cite{lovis2017}. The basic premise is to cross correlate high resolution spectra of exoplanet systems with known spectroscopic templates at different line of sight velocities. The planet's spectral features will be Doppler shifted relative to the stellar light and any Earth based atmospheric spectral lines and hence be distinguishable as a peak in the CC function (CCF) at the planet's line of sight velocity. Highly structured spectral features observed at very high spectral resolutions show many lines,  enabling detectable correlation signals when cross correlated with spectral templates. The high spectral resolution CC technique is becoming common in exoplanet science to probe both transiting and non-transiting hot Jupiter atmospheres \citep{birkby2018}. In Hot Jupiters, the technique has been used to detect the molecules CO and H$_2$O (in transmission and emission) as well as TiO and evidence of HCN in emission \citep{snellen2010, brogi2012, birkby2013, rodler2013,nugroho2017, hawker2018, cabot2019}.  Additionally, atomic species have been detected (Fe, Fe$^{+}$, Ti$^{+}$, \citep{2018Natur.560..453H}). The technique has also been used in combination with integral field spectroscopy to make molecular detections of CO and H$_2$O in emission for young self-luminous giant planets \citep{snellen2014, hoeij2018,delaroche2018,wang2018}. \\\\Recently, other studies explore looking for the O$_2$ A-band in transmission for nearby transiting planets around M dwarfs \citep{snellen2013, rodler2014} and also suggest novel instrumentation concepts for obtaining extremely high resolution spectra \citep{benami2018} . Whilst of great importance, performing such searches in transmission will likely be limited by the low occurrence rate of close-by transiting rocky planets  \citep{dressing2015}. In this work we simulate using the cross correlation technique in combination with high contrast imaging to search for the O$_2$ A-band in reflected light. Such a search would complement transmission studies well, given that the occurrence rate of non-transiting close-by rocky planets is higher and in Proxima b we already have the ideal candidate for this method  \citep{anglada2016}. We explore the technique by simulating observed stellar and reflected planet spectra for various point spread functions (PSF) and subtractions of stellar light which we cross correlate with spectral templates -- either the reflected stellar spectrum to detect a planet (assuming grey reflection) or O$ _2$ A-band spectral feature to characterise whether O$ _2 $ is present.\\\\
 Whether O$_2$ is likely to be present or survived in the atmosphere of Proxima b is the subject of many studies \citep{ribas2016, dong2017}. Even if it were present it has been suggested that both biotic and abiotic sources could explain such an oxygenated atmosphere potentially undermining the use of O$_2$ as a biosignature \citep{meadows2017,meadows2018} . On the other hand there can be no doubt that the discovery of an oxygenated atmosphere on an Earth-sized temperate zone exoplanet would demand extensive follow up, being a major discovery in the quest to discover extraterrestrial life. A truly convincing biosignature detection will require observing multiple chemical species and extensive characterisation of the atmosphere and other planetary and stellar properties \citep{schwieterman2018}.
\\\\We simulate observations of a  Proxima b-like planet and the O$_2$ A-band spectral feature using the ELT and its proposed high resolution spectrographs HIRES ($ R\approx150,000 $,  \citet{hires2016}) and HARMONI ($ R\approx20,000 $,  \citet{harmoni2014}). We use a generalised analysis to understand the requirements on the PSFs of instruments and telescopes utilising the CC technique as well as the systematic errors in subtracting the stellar light from observed spectra. Section 2 describes the methods in simulating IFS data and performing the CC analyses. We present the results of various simulations for Proxima b in section 3. In section 4 the analysis is generalised further to the ELT for detecting and characterising other exoplanet systems with various system distances, stellar spectral types, planetary radii and orbital separations.  We present our conclusions in section 5.

\section{Methods}
\subsection{Simulating IFS Data}
\begin{figure}
	% To include a figure from a file named example.*
	% Allowable file formats are eps or ps if compiling using latex
	% or pdf, png, jpg if compiling using pdflatex
	\includegraphics[width=\columnwidth]{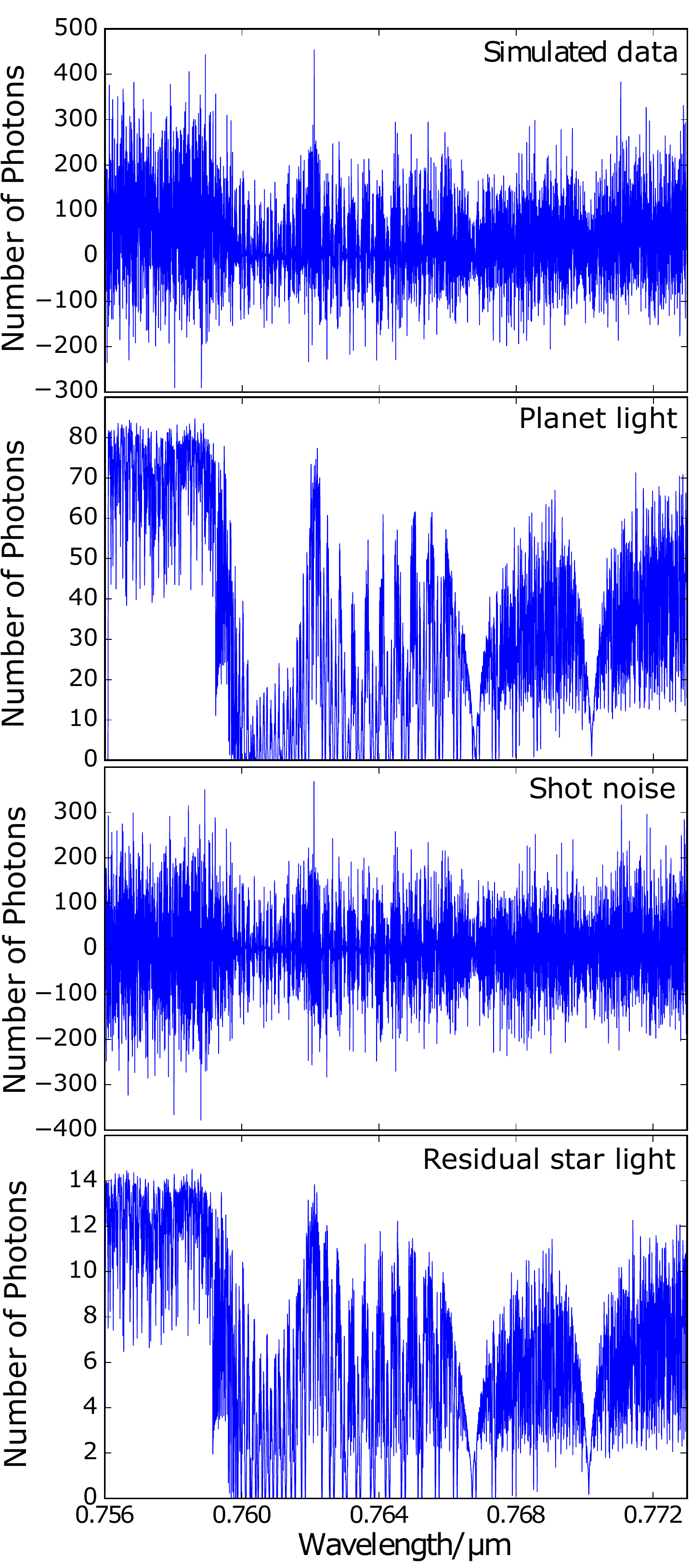}
    \caption{The simulated data at the planet spaxel after stellar subtraction (top). The three components of the simulated data (planet light, shot noise and residual star light) are also plotted. Note the residual starlight is that left over from HCI and after subtraction, parametrised by $C_i$ and $I$ respectively. The shot noise is dominated by the counts in starlight prior to subtraction. This example case has $C_i$=10$^{-4}$, $I=10^{-3}$, $TC_v=10^{-1.5}$ hours giving a  planet spectrum signal-to-noise ratio, SNR$_{spec}$ = 0.55.}  
    \label{fig:components of data}
\end{figure}

We simulated integral field spectrograph (IFS) observations of the Proxima Centauri system in the visible for high spectral resolutions up to R$\sim$150,000 and various PSFs. In all cases we modelled the stellar light using the high resolution spectrum from the PHOENIX library \citep{2013A&A...553A...6H} with the closest parameters to those of Proxima Centauri ($ T_{eff}=3000$ K, [Fe/H]$ = 0$ dex and $ \log_{10}(g)=5$). This stellar spectrum was Doppler shifted by the star's barycentric line of sight radial velocity (RV).  The rotation of the Earth (<0.5kms$^{-1}$) was not included and is negligible for the spectral resolutions considered.\\\\
The planet light was modelled in two ways, either as simply the stellar light times the planet-star contrast $ C_{pl}$ (grey reflection, no atmosphere) or as before but with a geometric albedo modelled as the transmittance of Earth's atmosphere (oxygenated case) following similar methods to \cite{lovis2017}. Using the transmittance of the Earth's atmosphere to approximate the geometric albedo will result in relative line depths different to the the true reflection spectrum of the Earth; however, given that such line depths will be strongly dependent on the atmospheric composition, pressure-temperature profile, clouds and hazes of Proxima b, the single-pass transmittance model is a reasonable approximation for a potential oxygenated atmosphere. The planet -- star contrast, $ C_{pl} $, was calculated according to equation (\ref{contrast}) assuming a grey albedo $a=0.3$, orbital radius $d_{pl}=0.032$ AU,  angular separation from the star of 25mas, planet radius of $ r_{pl}=1.5 R_{\oplus}$, and where  $\Phi(\alpha)$ is the phase law from \citet{2010exop.book..111T}. These parameters are the same as those in \cite{2015A&A...576A..59S} and likewise we use the maximum of the phase law  $\Phi(\alpha)=0.5$, which gives $ C_{pl} = 6\times10^{-7} $. We modelled the orbital motion of the planet by Doppler shifting the spectrum by a further 30 km s$^{-1}$. Both the planet and stellar light are subsequently multiplied by the transmittance spectrum of the Earth's atmosphere ($t_E$) to account for the observations being made from the ground.

\begin{equation}
C_{pl}=a\bigg(\frac{r_{pl}^2}{d_{pl}^2}\bigg)\Phi(\alpha)=a\bigg(\frac{r_{pl}^2}{d_{pl}^2}\bigg)\frac{1}{\pi}\big(\sin(\alpha)+(\pi-\alpha)\cos(\alpha)\big)
\label{contrast}
\end{equation}
We simulated IFS data at a given spaxel for various PSFs, exposure times ($T$) and subtractions of the stellar light using equation \ref{dataequation}.

\begin{equation}
\label{dataequation}
\text{data}=TC_vt_E(C_{pl}\hat{S}_{pl} + C_iI)f_{s} + \mathcal{N}(0,1)\sqrt{TC_vt_E(2C_i+C_{pl}\hat{S}_{pl})f_{s}}
\end{equation}
The PSFs were parametrised by $C_i$, the instrument contrast, and $C_v$, the central value of the normalised PSF. $C_i$ is defined across the field of view as the ratio of the stellar light at a given position to the stellar light at the star position. $I$ parametrises the systematic error in subtraction of the stellar light and is defined as the fraction of stellar light that remains. $\hat{S}_{pl}$ represents the reflection of stellar light by the planet, encoding both the geometric albedo and the Doppler shift. Lastly $f_s$ is the number of stellar photons incident on the telescope detector per hour assuming a collecting area of the ELT , 976m$^2$, and total throughput of 15\% as in \cite{2015A&A...576A..59S}. The simulated data has three components: the planet light, residual stellar light and photon shot noise. Only the photon shot noise from the planet signal and stellar subtraction is considered as this is the fundamental limitation -- readout, background and dark current noise sources are neglected. Figure \ref{fig:components of data} shows the simulated data and each of its individual components (planet light, shot noise and residual star light) for an example Earth-like case with $C_i=10^{-4}$, $I=10^{-3}$, $TC_v  = 10^{-1.5}$ hours. A useful metric for the quality of the simulated data is the signal-to-noise ratio  of the planet spectrum ( SNR$_{spec}$) averaged over wavelength. It is defined in equation \ref{SNR} as the ratio of the planet light to the shot noise.

\begin{equation}
\text{SNR}_{spec}=\sqrt{\frac{TC_vf_{s}}{2C_i}}C_{pl}=\sqrt{\frac{TC_vf_{s}}{2C_i}}a\bigg(\frac{r_{pl}^2}{d_{pl}^2}\bigg)\Phi(\alpha)
\label{SNR}
\end{equation}
\subsection{Cross-correlation analysis}
\begin{figure}
	% To include a figure from a file named example.*
	% Allowable file formats are eps or ps if compiling using latex
	% or pdf, png, jpg if compiling using pdflatex
	\includegraphics[width=\columnwidth]{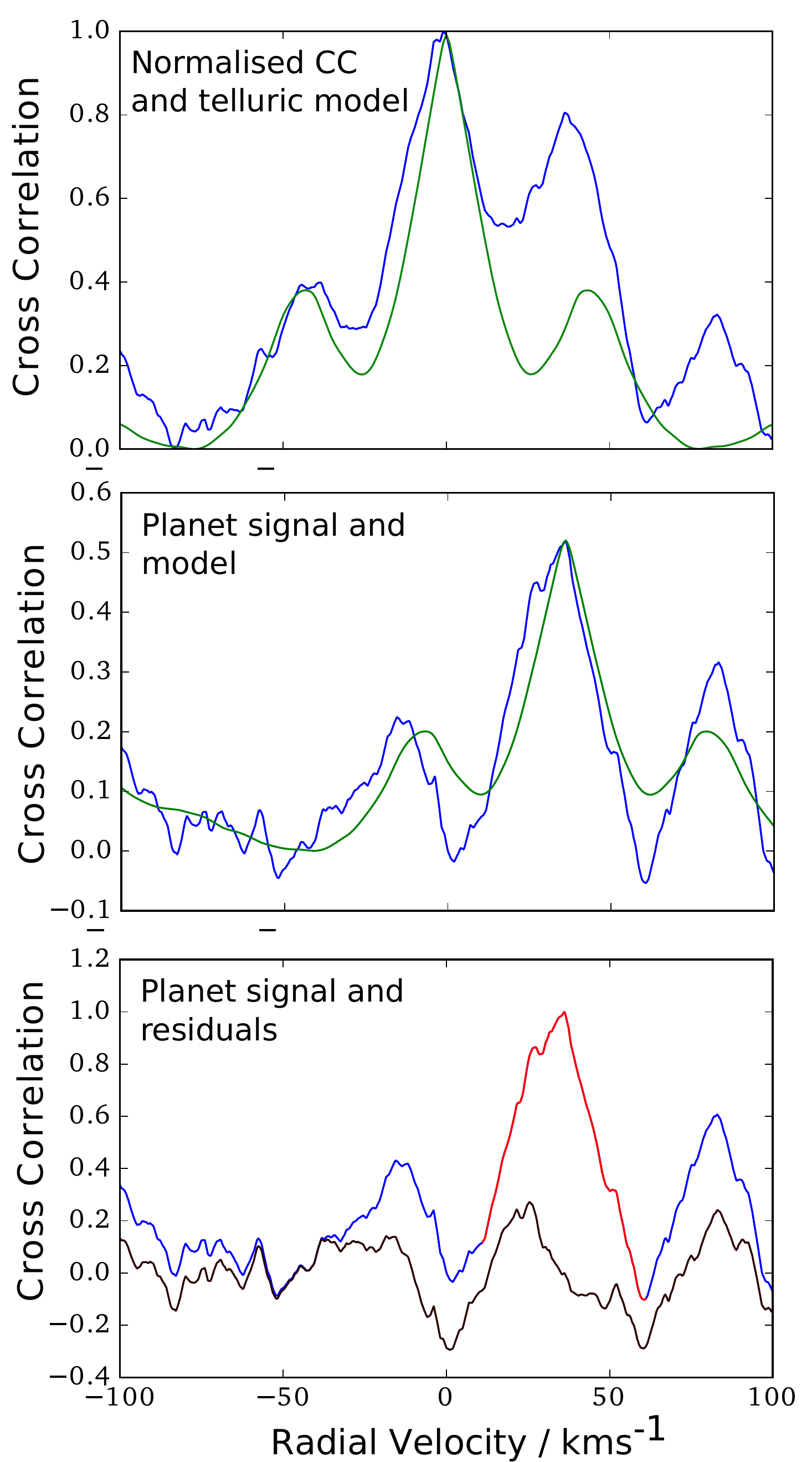}

    \caption{The stages of CC analysis for detecting the O$_2$ A-band. The CCFs are shown (blue) along with  model CCFs (green) which are based on the autocorrelation of the O$_2$ spectral template. The top panel shows the raw CCF and telluric model which is subtracted off to give the planet signal in the middle panel. The planet signal is then renormalised and the residuals found by subtracting off the planet model. The ratio of the variances of the highlighted normalized planet peak signal (red) and the residuals (brown) is used as the  SNR$_{CC}$ metric for the signal-to-noise in the CCF. This example case uses $C_i$=10$^{-4}$, $I=10^{-3}$, $TC_v=10^{-1.5}$ hours giving  SNR$_{CC}$ = 6.56.  }
    \label{fig:components of correlation}
\end{figure}

We perform CC analyses with various planet templates on the simulated data using the CC function, \texttt{pyasl.crosscorrRV}, in the \texttt{PyAstronomy} package. In the case of \emph{planet detection}, we cross-correlate the simulated data with the model stellar spectrum and look for a detectable signal at the planetary RV and spatial position in the FOV. In the case of \emph{planet characterisation}, we cross correlate the simulated data with the O$_2$ A-band spectral feature from the telluric transmission spectrum. Again we look for a detectable signal at the planetary RV and spatial position in the FOV, however we now ignore the signal at 0 km s$^{-1}$ due to the absorption of the Earth's atmosphere. \\\\ In the case of planet characterisation stellar lines are not included in the template used for CC. Their omission may slightly underestimate the detection significance however including them risks misleadingly boosting the significance of an O$_2$ detection. This is because the CC is sensitive to the lines that are present and the ratio of the line depths. Not including stellar lines in the template means that the O$_2$ relative line depths are not as well matched between the template and the simulated data as they would be if we convolved the O$_2$ template with the stellar spectra. However if included in the template the stellar lines could correlate well with those stellar lines in data irrespective of whether it contains the spectral of signature O$_2$ and erroneously boost the O$_2$ detection significance. Hence though potentially pessimistic, looking for O$_2$ in isolation is very robust. \\\\
A useful metric for the CC detection significances is the ratio of the variances of the  Gaussian-like detection feature and noise in the CC function, this gives the CC signal-to-noise ratio ( SNR$_{CC}$). In calculating the  SNR$_{CC}$ for O$_2$ A-band detections, care must be taken to subtract off the CC signal from the Earth's atmosphere and account for the CC structure inherent to the O$_2$ A-band feature as seen from its autocorrelation. In particular, due to the many line doublets in the O$_2$ A-band feature, when correlated with itself minor peaks result $\pm45$ km s$^{-1}$ either side of the main detection peak. The following method is used to obtain the  SNR$_{CC}$ of O$_2$ CC detections (for more details see figure \ref{fig:components of correlation}):
\begin{itemize}
	\item[1.]The CCF is normalised and the telluric 0 kms$ ^{-1} $ feature is modelled as the autocorrelation of the O$ _2 $ feature and subtracted off.
	\item[2.]The remaining planet signal is then normalised and modelled as the autocorrelation of the O$_2$ feature.
	\item[3.]An estimate of the noise is obtained from the residuals between the normalised planet signal and planet model.
	\item[4.] We found the ratio of the variances of the planet signal and the residuals, ( SNR$_{CC}$), is a useful metric for determining the significance of a detection.
\end{itemize}

\section{Simulations}

We used the method to conduct simulations for various PSFs and spectral resolutions outlined below. First, we checked our code and method by reproducing the results of \cite{2015A&A...576A..59S}, detecting Proxima b simply using the reflected stellar light with 10 hours of exposure time. We then focussed on using the method to simulate the detection of the O$_2$ A-band in Proxima b in reflected light. Subsequent simulations then explore the dependence of the HCI+HRS method to detect the O$_2$ A-band upon the instrument contrast and stellar systematics by simulating the data at the planet spaxel.

\subsection{Detection of Proxima b by CC}\label{3.1}

\begin{figure}
	% To include a figure from a file named example.*
	% Allowable file formats are eps or ps if compiling using latex
	% or pdf, png, jpg if compiling using pdflatex
	\includegraphics[width=\columnwidth]{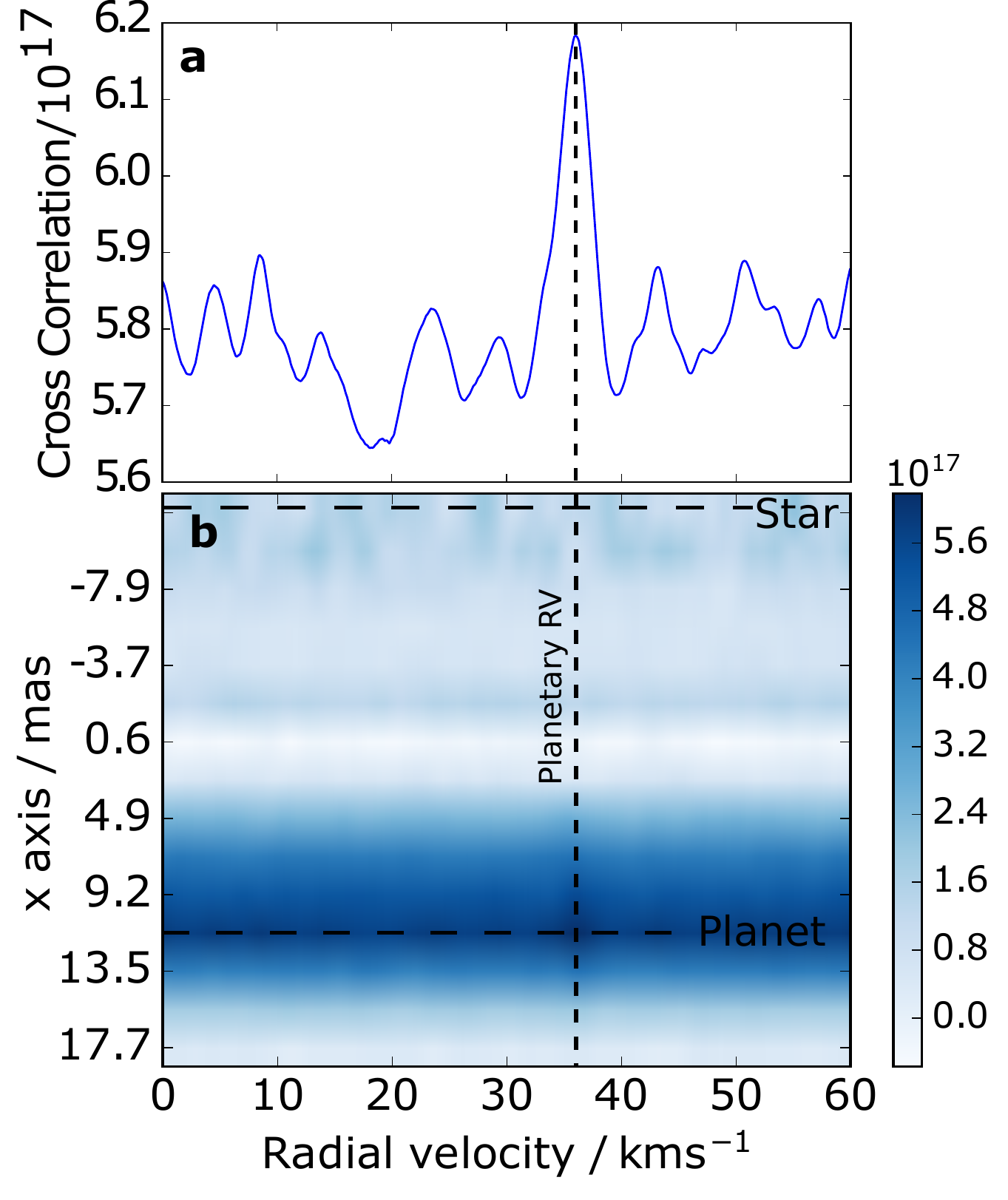}
    \caption{The detection of  a Proxima b-like planet using the reflected stellar light. Panel (a) shows the CCF at the planet spaxel, note the significant peak at the expected RV of 36kms$^{-1}$. Panel (b) shows the CCF in the y=0 plane (where the y-axis runs through the star and planet) of the CC data cube, note the strong band of correlation at the planet position (x=12.5 mas) as the planet spaxel is brighter than the surrounding spaxels and the peak correlation at the expected RV as confirmed by the cross section shown in panel (a).  An exposure time of 10 hours is assumed with an orbital phase of 0.5 and perfect subtraction of stellar light.}
    \label{fig:planet detection}
\end{figure}

We simulate the PSF of the ELT following the methods of previous work \citep{2015A&A...576A..59S}. Using the POPPY package \citep{2016ascl.soft02018P}, we generated a diffraction limited PSF, PSF$_{DL}$, for the ELT aperture. This PSF was modified to account for atmospheric seeing using a Strehl ratio of 30\% and adding a normalised seeing-limited halo with a relative strength of 70\%. The PSF is described by equation (\ref{seeinglimitedPSFeq}) where $ \alpha=265.4 $, $ \beta=2.5 $ (note this is a Moffat profile)  and $N_0$ is a normalisation factor. The PSFs were then cropped to give an FOV of 0.6" by 0.6"  -- the typical FOV needed for resolving habitable zone, nearby, M-dwarf planets.
\begin{equation}
	\text{PSF} = 0.3\text{ PSF}_{DL} + 0.7 N_{0}(1+(r/\alpha)^2)^{-\beta}
	\label{seeinglimitedPSFeq}
\end{equation}
We used this PSF with the methods described above to generate a data cube with spatial information given by 1st and 2nd dimensions; the 3rd dimension gives the values containing the spectral information as would be produced by an IFS. The wavelength coverage is 0.735 $\mu$m to 0.785 $\mu$m. Here we assume perfect stellar subtraction ($I=0$), i.e. the stellar light is reduced to shot noise and an exposure time of 10 hours. The planet light is modelled with a grey reflection and no atmosphere. We cross correlate the simulated data with the stellar template and detect the planet in the CCF at the expected RV = 36 km s$^{-1}$ with a  SNR$_{CC}$ of 10 - in good agreement with the results of \citep{2015A&A...576A..59S}. The detection is shown in figure \ref{fig:planet detection}; panel (a) shows the CCF for the planet spaxel with a clear detection at 36kms$^{-1}$ where the CCF is greatest. Note in panel (b) (the y=0 plane of the CCF cube) there is a strong dark band at the planet position for all RVs and panel (a) is simply the cross section  along this dark band.  Effectively the dark band in (b) indicates that the planet spaxel is brighter than the surrounding spaxels but as shown in (a) the spectral lines are only best correlated at the correct planetary RV. Since the cross correlation is the sliding dot product with RV, the absolute cross correlation values plotted here are sensitive to the continuum level at each spaxel. The stellar signal has been perfectly subtracted thus what remains away from the planet position is shot noise with a continuum level $\approx$0. At the planet position the reflected stellar signal has a higher continuum meaning at these spaxels the baseline of the CCF is much higher (around $5.8\times 10^{17}$ as seen in panel a) and the CC (sliding dot product) does not go to zero but is maximum at the planetary RV. The structure of the PSF is also reflected in similar fainter bands at all RVs next to the planet position. The star is not seen, as this is the idealised case where stellar light is perfectly subtracted. 

\subsection{Detection of O$_2$ in Proxima b by CC}\label{3.2}
\begin{figure}
	% To include a figure from a file named example.*
	% Allowable file formats are eps or ps if compiling using latex
	% or pdf, png, jpg if compiling using pdflatex
	\includegraphics[width=\columnwidth]{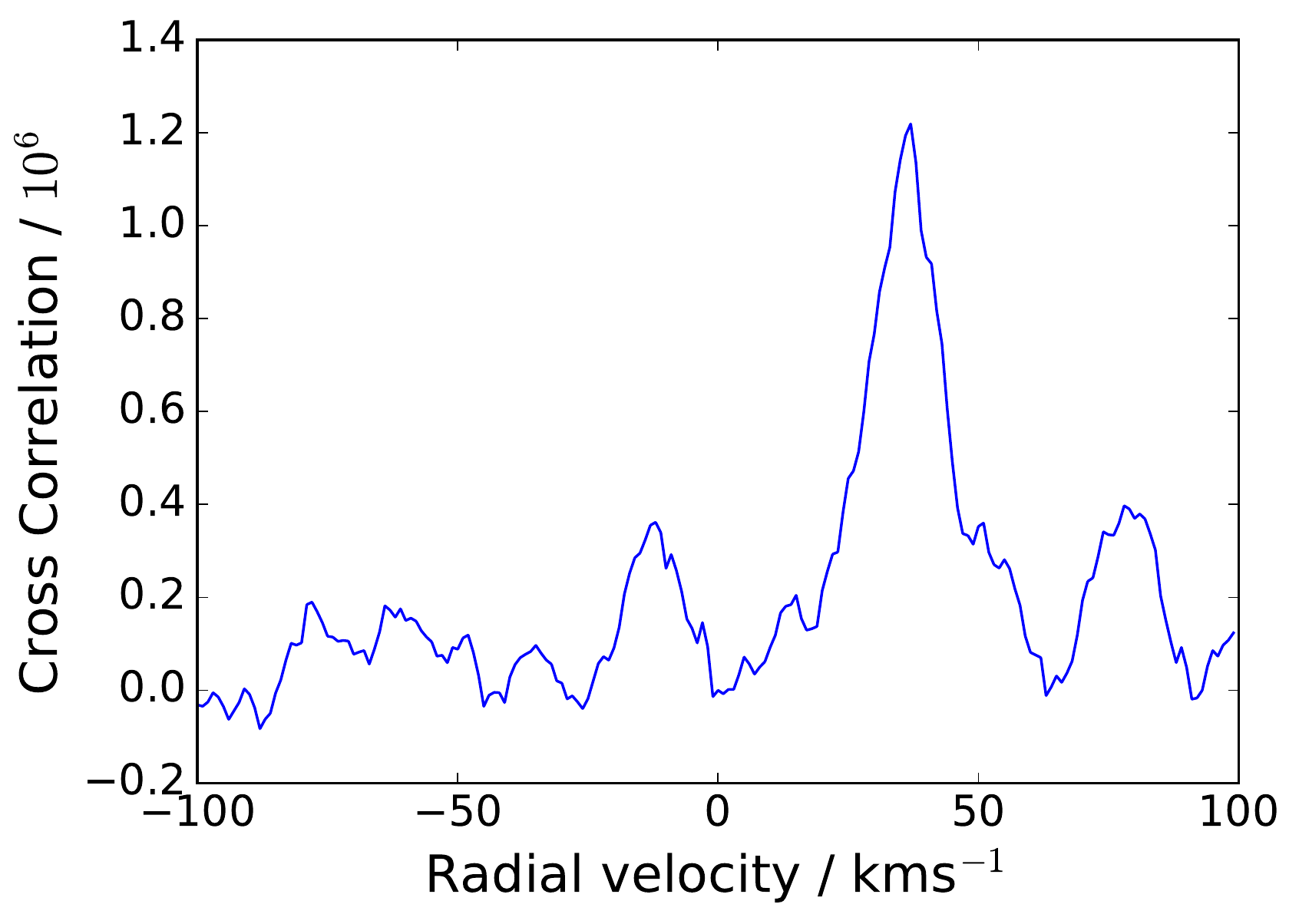}
    \caption{The CCF at the planet spaxel for CC  with the O$_2$ A-band template for an exposure time of 70 hours. Note the significant peak at the expected RV of 36kms$^{-1}$ which constitutes a >5$\sigma$ detection. }
    \label{fig:det O_2}
\end{figure}

We use the PSF from the previous section to simulate the detection of the O$_2$ A-band in reflected light for Proxima b. We now model the planet light using the Earth-like case described in the methods section and use a wavelength coverage of 0.756 $\mu$m to 0.773 $\mu$m . We find that a minimum exposure time between 30 to 70 hours is required to detect the O$_2$ A-band  with SNR$_{CC}$ ranging from 3 to 5. There is variation in the SNR$_{CC}$ obtained for identical exposure times due to the variation seen in repeating the simulations due to the variation of shot noise with the normal distribution factor. Repeated simulations show consistent detections in the CCF with  SNR$_{CC}$ $\geq 5$ for $T\geq70$ hours.  Note this is assuming a constant orbital phase of 0.5 which would require many observations over a long time given the planet's orbital period of 11.2 days. An example CCF for 70 hours is shown in figure \ref{fig:det O_2}, note here the atmospheric absorption CCF feature has been subtracted off using the methods detailed above. We also tested for false positives by CC of the data with the O$_2$ template when the planet light was modelled as the grey reflection and no atmosphere case. We see no significant correlation at the planetary RV in this case but only  structure intrinsic to the autocorrelation of the telluric O$_2$ feature centred at 0 kms$^{-1}$ with its minor peaks $\pm45$ km s$^{-1}$ due to the many line doublets in the O$_2$ A-band feature.

\subsection{Investigating dependence on $C_i$, $I$ and $R$}
The PSF used above is likely over optimistic, with a Strehl ratio of 0.3 being difficult to achieve for wavelengths below 1$\mu$m. We conducted further simulations investigating the dependence of detecting the O$_2$ A-band feature on both the instrument contrast ($C_i$) at the planet spaxel and the systematic error in stellar light subtraction, $I$. In these simulations we simulated data at the planet spaxel using equation \ref{dataequation} for a wide range of $C_i$ and $I$. We also considered a range of resolving powers 20,000 $\leq R \leq$150,000, with the upper limit corresponding to the resolution of the proposed HIRES instrument \cite{} and the lower limit to that of the first light HARMONI spectrograph \cite{}. In the following simulations we first investigated the shot noise limited case, followed by the systematics limited case.

\subsubsection{Shot noise limited case}
\begin{figure}
	% To include a figure from a file named example.*
	% Allowable file formats are eps or ps if compiling using latex
	% or pdf, png, jpg if compiling using pdflatex
    \includegraphics[width=0.97\columnwidth]{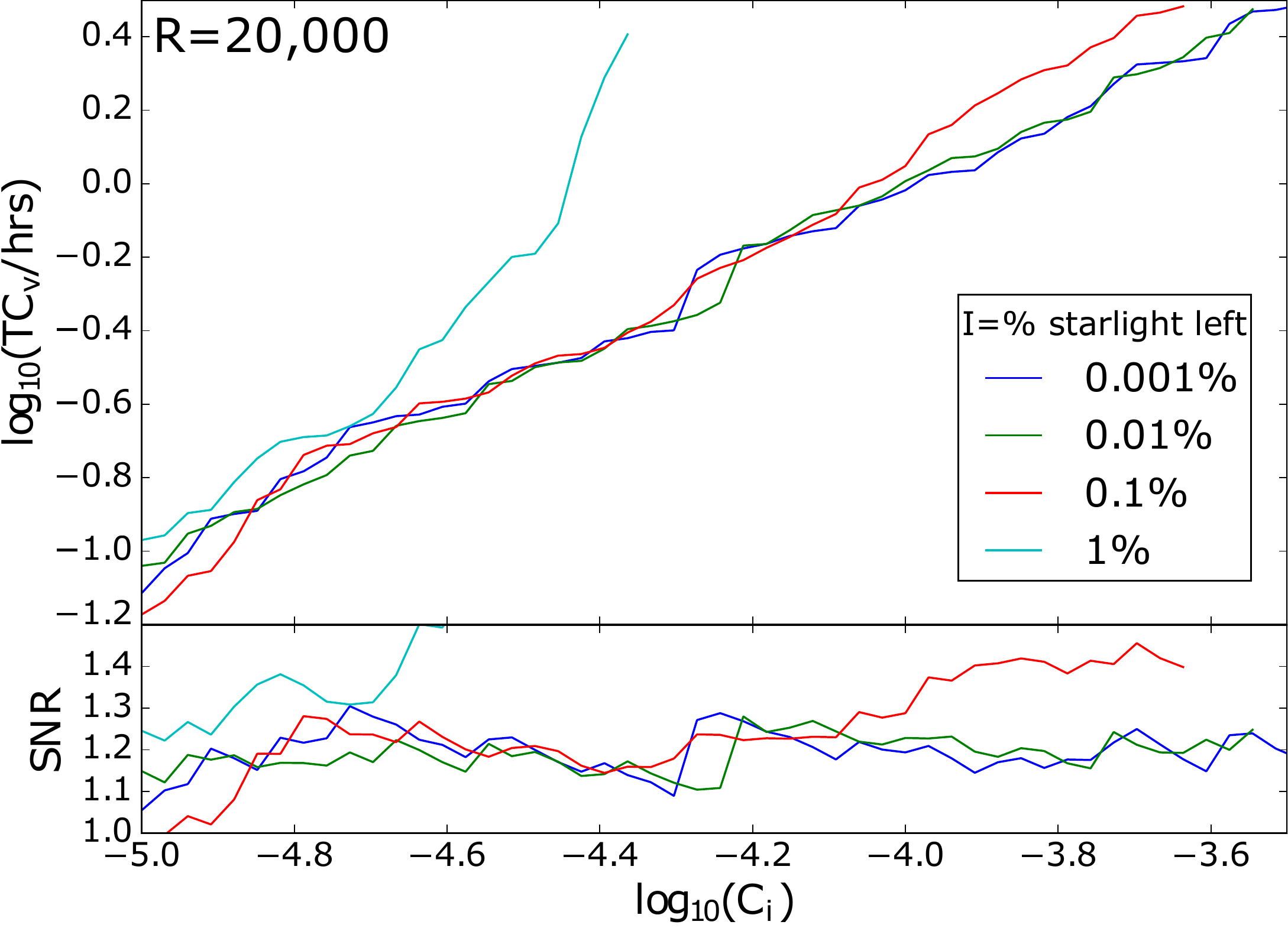}
    \includegraphics[width=\columnwidth]{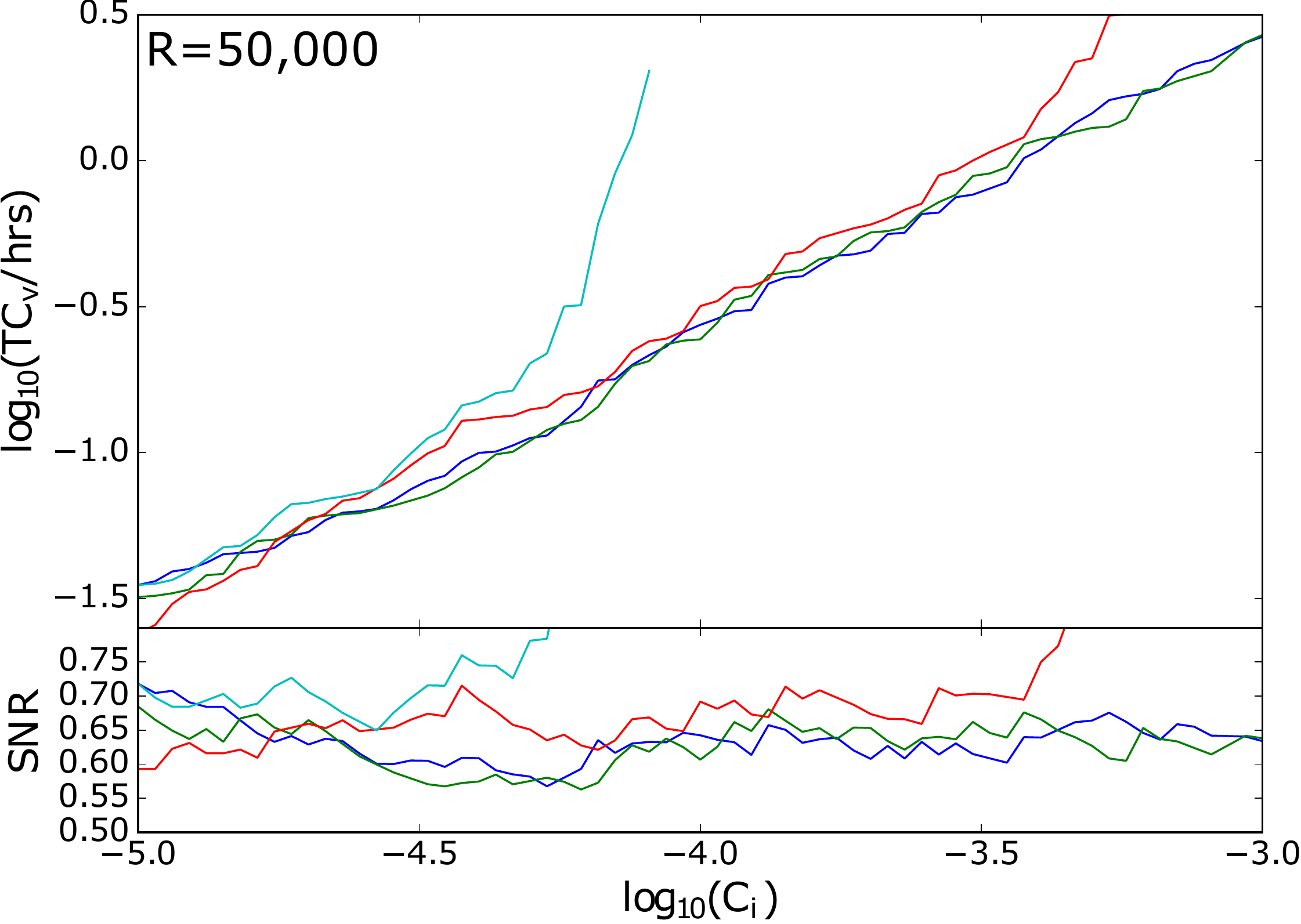}
	\includegraphics[width=\columnwidth]{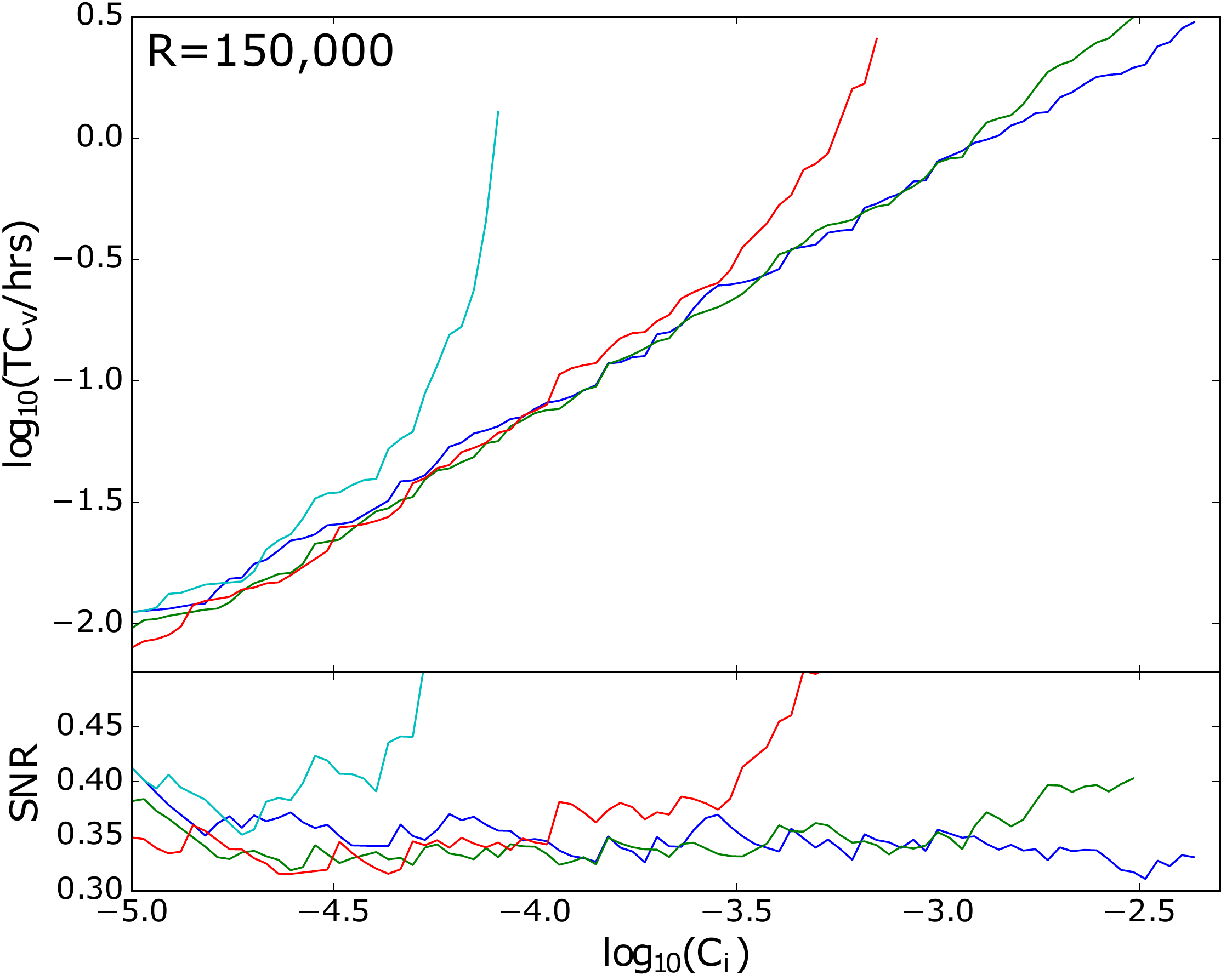}
\caption{ SNR$_{spec}$ and $TC_v$ required for detecting O$_2$ against the instrument contrast for a range of $I$ and resolving powers. For noise limited cases these give threshold SNR$_{det}$ (for detecting the O$_2$ A-band) of 0.35, 0.65 and 1.2 for resolving powers of 150,000, 50,000 and 20,000 respectively.}
    \label{fig:Ci vs TCv}
\end{figure}

We ran simulations for a range of $ I $ and $ C_i $ determining the minimum exposure time required for an ELT sized collecting area = 976m$ ^2 $. We trial each exposure time 10 times and find the minimum exposure time that returns a CC detection with  SNR$_{CC}\geq3$ at least 9 times out of the 10 trials. The multiple trials account for the variation due to the random draws taken from a normal distribution in modeling the shot noise. We then estimated the required  SNR$_{spec}$ for an O$_2 $ A-band CC detection for the given $ I $ and $ C_i $ using equation \ref{SNR}. This required  SNR$_{spec}$ of the simulated data is a general result applicable to ground-based observations with different collecting areas, $C_V$ and throughputs. The following simulations use $ C_v $=0.016  the central value from the PSF used in section 3.1 and $ C_{pl}=6\times10^{-7} $. The results are shown for various resolving powers in figure \ref{fig:Ci vs TCv} and discussed below.\\\\
The $\log_{10}(TC_V)$ sections of the figure give the minimum exposure time and accuracy of subtraction needed assuming a PSF with some $C_v $, $C _i $ for a collecting area of 976m$ ^2 $. For example with $C_v $=0.01, $C_i $=10$ ^{-3.8} $ then a minimum exposure time of around 10 hours is needed with a subtraction within at least 0.1\% for $R$=150,000. There are two parts to each of these curves:\\\\
1. At poorer contrasts there is a cut-off contrast for a given $I$ where the starlight component, from systematic errors in the subtraction, overwhelms the planet light and increased T no longer improves chances of detection (see section 3.3.2). \\\\
2. For better contrasts than the cut-off contrast there is a linear section with a slope $ \approx $ 1, indicating a roughly constant  SNR$_{spec}$.\\\\
The  SNR$_{spec}$ sections of the figure give estimates of the minimum (or threshold)  SNR$_{spec}$, $\text{SNR}_{det}$, needed to make a detection for a given $ C_i $ and $ I $. These show the  $\text{SNR}_{det}$ for a detection is approximately constant until the cut-off contrast is reached. Note also as R decreases the  $\text{SNR}_{det}$ required increases. The $\text{SNR}_{det}$ determined in these figures ranges from 0.35 to 1.2 for R from 150,000 to 20,000. 
\subsubsection{Systematics limited case}
\begin{figure}
	% To include a figure from a file named example.*
	% Allowable file formats are eps or ps if compiling using latex
	% or pdf, png, jpg if compiling using pdflatex
%	\includegraphics[width=\columnwidth]{I_vs_Ci_EELT_R_150000.pdf}
%    \includegraphics[width=\columnwidth]{E-ELT_Ci_vs_TCv_and_SNR_R_125000.pdf}
%	\includegraphics[width=\columnwidth]{I_vs_Ci_EELT_R_100000.pdf}
%	\includegraphics[width=\columnwidth]{I_vs_Ci_EELT_R_50000.pdf}
	\includegraphics[width=\columnwidth]{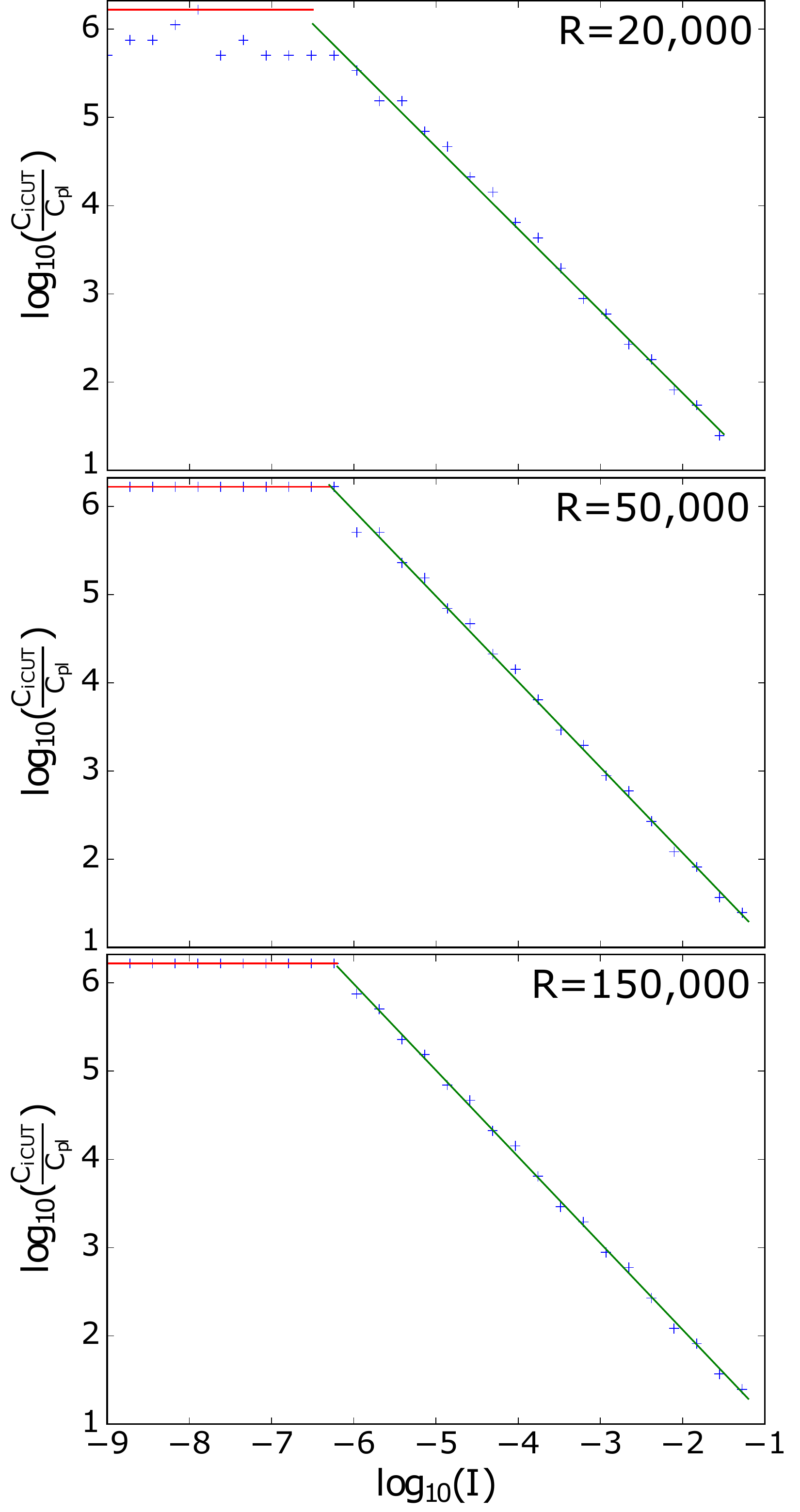}

    \caption{The cutoff instrument contrast, $C_{iCUT}$ normalised by the star-planet contrast against $I$, the fraction of stellar light remaining, for various resolving powers. Linear regression analysis of the results (blue crosses) give: $m=0.98\pm0.01, c=0.11\pm0.04$ for R=150,000, $m=0.97\pm0.01, c=0.13\pm0.04$ for R=50,000, $m=0.93\pm0.01, c=0.016\pm0.06$ for R=20,000 with these lines plotted in green. The red line indicates the limit where $ C_i=1 $ i.e. the planet is spatially unresolved. }
    \label{fig:I vs Ci}
\end{figure}

As noted above, we found there is a cut-off instrument contrast, $ C_{iCUT} $ for a given $ I $ and $ C_{pl} $ beyond which increasing $ TC_v $ no longer increases the chance of detection. This is due to systematic errors in the subtraction of the stellar light overwhelming the planet light. We investigated how systematics effect our ability to make detections by testing combinations of $ I $ and $ C_i $ for various R with an exposure time of 500,000 hours to determine the dependence of $ C_{iCUT} $ on $I$ for different resolving powers.  This extreme exposure time is used in order to find the asymptotic limit where increased exposure time no longer improves detections because of systematic errors in the subtraction of stellar light are dominant. The results were plotted with $ \log_{10}(C_i/C_{pl}) $ against $ \log_{10}(I) $ with a linear fit:

\begin{equation}
\log_{10}\Bigg(\frac{C_{iCUT}}{C_{pl}}\Bigg)=-m\log_{10}(I)+c
\end{equation}
Given a good linear fit and for gradients $ m\approx1 $ we find:
\begin{equation}
\frac{\text{remaining starlight}}{\text{planet light}} = \frac{IC_{iCUT}}{C_{pl}}\approx10^{c}
\label{systematics equation}
\end{equation}
 \\See figure \ref{fig:I vs Ci} for the results (blue crosses), where m and c are determined by linear regression of the section with non-zero gradient resulting in the green line. The red line indicates the limit where $ C_i=1 $ i.e. the planet is spatially unresolved. Note the R=50,000+ cases agree on a maximum $ IC_{iCUT}/C_{pl}\approx1.3 $ to one $ \sigma $ with the R=20,000 results giving a slightly reduced value. This suggests increases in R only marginally reduce the effect of systematics, though note the $ m\approx1 $ approximation is worse for lower R. We can use these relationships to obtain the cut-off contrast where systematics dominate for a given $ I $, $ C_{pl} $ and R. For example with R=150,000, we see if $ I=10^{-5}$ and $ C_{pl}=10^{-7} $ then $ C_{iCUT}=10^{-2}$. 
 
\subsection{Observing with possible ELT PSFs}
\begin{figure}
	\centering
	\includegraphics[width=\columnwidth]{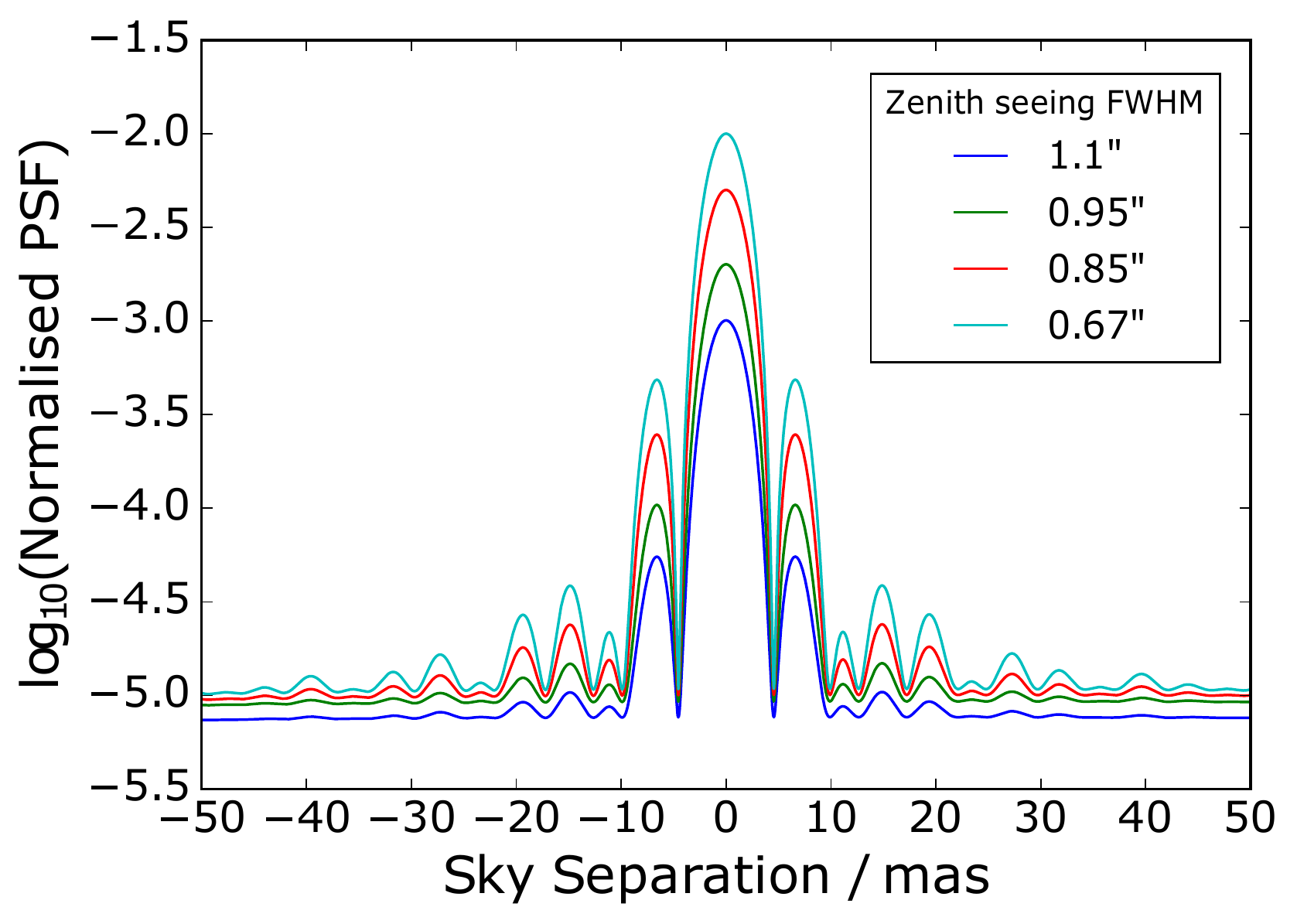}
	\caption{The ELT PSF at 0.75$ \mu $m using model parameters from \citet{thatte} for various zenith seeing full width half maxima (FWHM).}
	\label{fig: ELT Ci vs r}
\end{figure}

We used the parametrisation described in \cite{thatte} to model the ELT PSF at 0.750$\mu$m and obtain better estimates of possible $ C_i $ values. The resulting PSF for various zenith seeing FWHM is shown in figure \ref{fig: ELT Ci vs r}. Note for Proxima b with sky separation of 25mas this gives $ C_i= 10^{-3}$ to $10^{-2.2} $ and $ C_v= 10^{-2}$ to $10^{-3}$. HIRES (R=150,000) would require the stellar light subtraction to be within 0.08\% to 0.01\%, and an SNR$ _{det} $ of 0.4 implies an exposure time between 55 and 3500 hours depending on the zenith seeing FWHM. The shorter exposure times are consistent with the results from optimistic PSFs in section \ref{3.2}. HARMONI, due to be a first light instrument, with a maximum R=20,000 would require the subtraction to be within 0.06\% to 0.0095\%,  SNR$ _{det} $=1.3, giving an estimated exposure time between 580 and 37,000 hours. Note the extremely strong dependence of exposure time on PSF  and the estimated exposure times assume an orbital phase of 0.5. Also these PSFs are without an extreme adaptive optics (XAO) based HCI system that would improve C$ _i$ values and should be present on any dedicated planet finding/characterising instrument to be coupled with a HRS spectrograph. Also we note, that the proposed design of the HARMONI spectrograph does not cover this wavelength range at its maximum resolution of R=20,000.  

\subsection{Investigating the cross correlation technique to detect Proxima b}
\begin{figure}
	\centering
    	\includegraphics[width=\columnwidth]{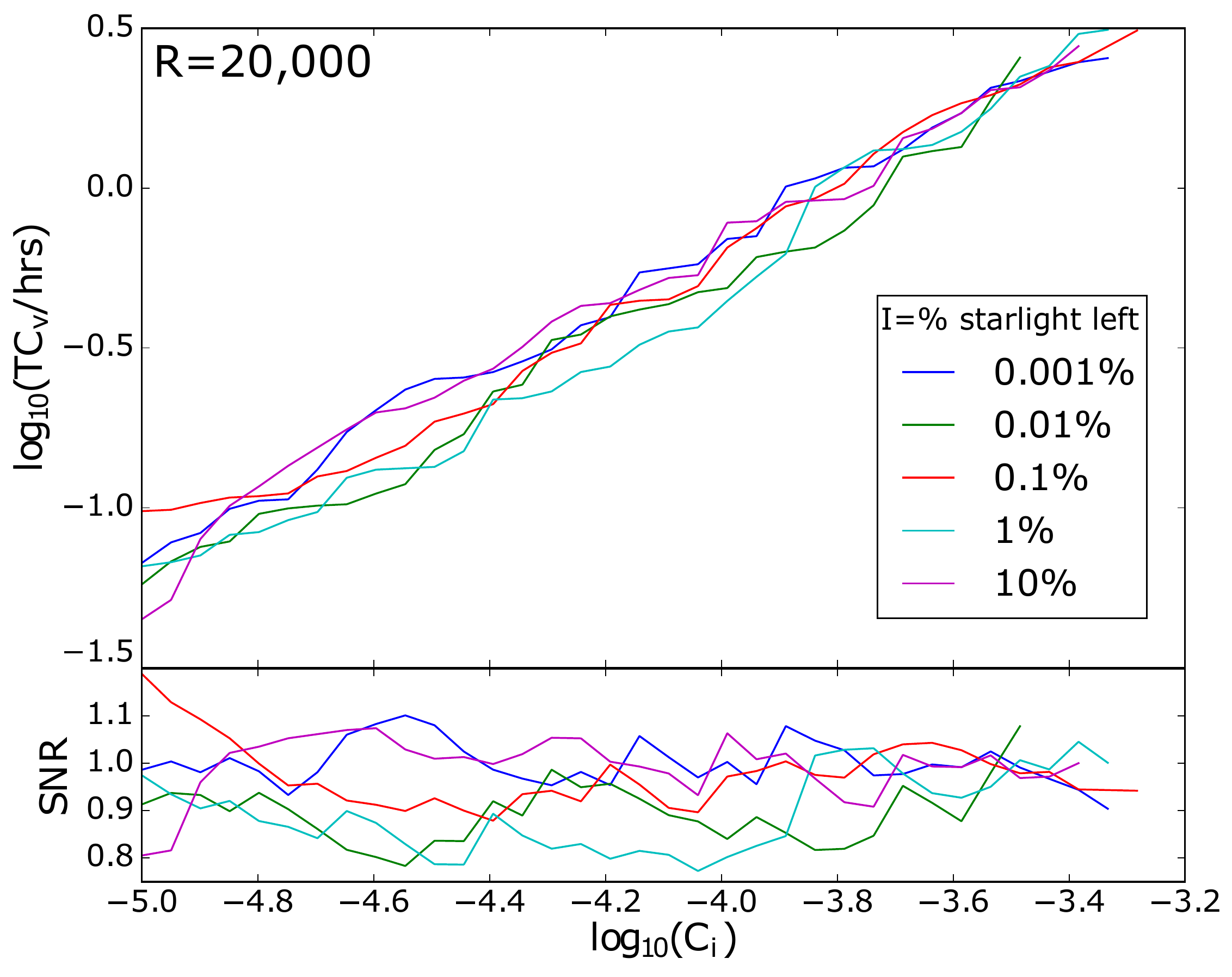}
		\includegraphics[width=\columnwidth]{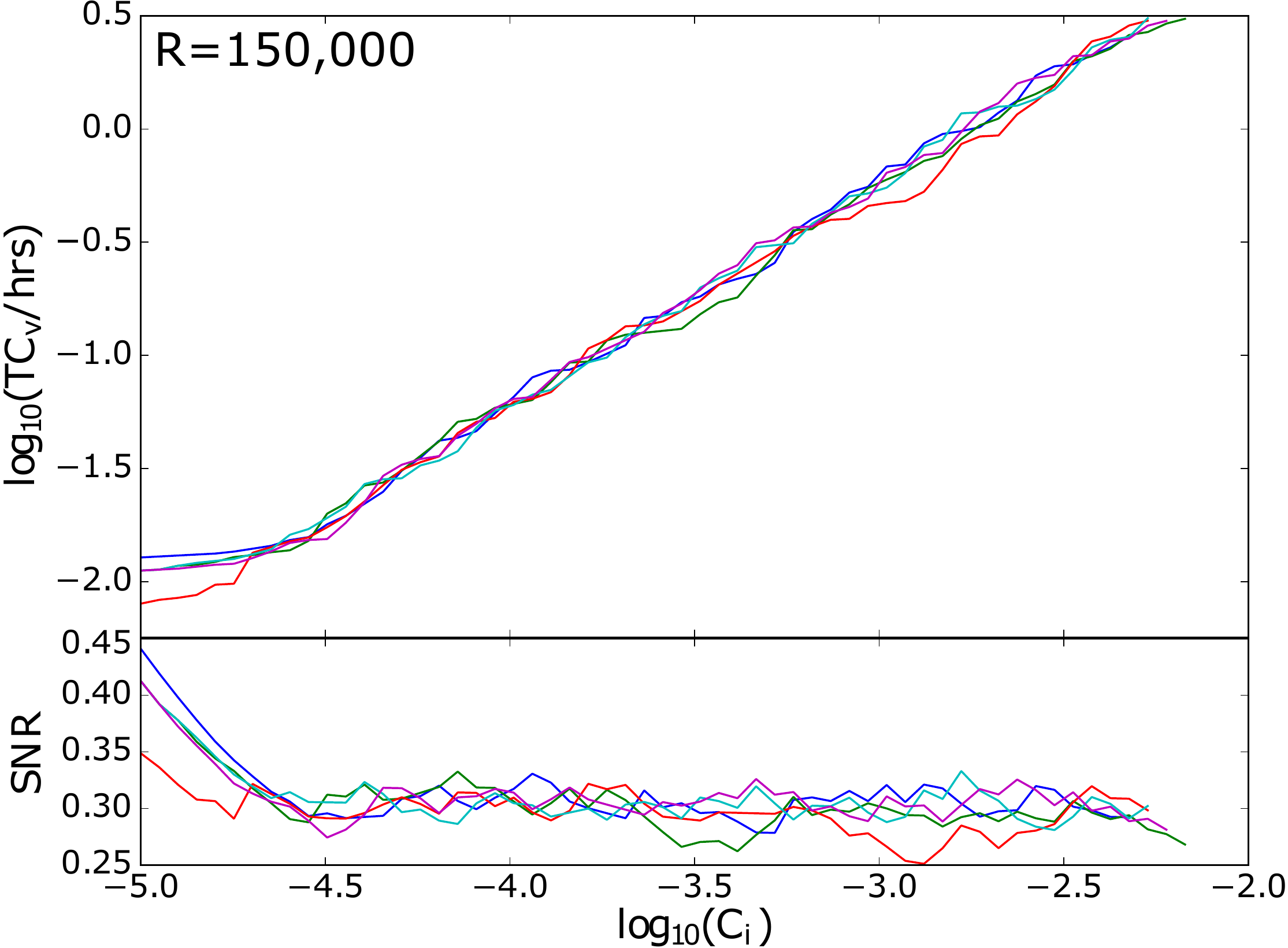}
		\caption{ SNR$_{spec}$ and $TC_v$ required for planet detection (grey reflection) against the instrument contrast, for a range of $I$ and resolving powers. For noise limited cases these give threshold SNR$_{det}$ (for detecting the the planet) of 0.3 and 1 for resolving powers of 150,000 and 20,000 respectively.}
		\label{Ci vs TCv detect}
\end{figure}

We repeat the above analysis of PSF dependence for observations aimed at the initial detection of the planet from the ground using the EELT . We generated the same simulated spectra and ran the same simulations except the CC analysis was carried out using the stellar spectrum as the template instead of the O$_2$ A-band. The template is used as if measured from the stellar light. This method depends on the stellar spectrum having a high degree of structure within the observed bandwidth; this is virtually always the case though the bandwidth could be extended if required. The results are shown in figure \ref{Ci vs TCv detect}, note we find no cutoff due to systematics for $I\leq 10\%$.
Given the model ELT PSFs (figure \ref{fig: ELT Ci vs r}), we suggest a detection could be made with HIRES for an exposure time of between 31 and 2000 hours and with HARMONI for between 350 and 22,000 hours, with little constraint from systematics due to CC with the stellar spectrum. Note again the extreme dependence on the PSF and hence the AO system.  These estimated exposure times also assume an orbital phase of 0.5. Although slightly more pessimistic,  the results for the best case PSFs are consistent with those from section \ref{3.1} and \cite{2015A&A...576A..59S} which use similarly optimistic PSFs. These results could potentially be improved by increasing the spectral range including more spectral lines to improve the CC -- the simulations used the bandwidth 0.756$\mu$m to 0.773$\mu$m.

\section{The observable parameter space with HRS CC
	}

Previous sections suggest that the  planet spectrum signal-to-noise SNR$_{spec}$ required to both detect exoplanets and characterise O$ _2$ (or other species with strong spectral features) in their atmospheres is significantly reduced by using CC of high resolution spectra. In determining the observable parameter space of the technique for finding and characterising exoplanets in reflected light (wavelength $ < $1$ \mu$m), we use equation \ref{SNRdet} for the threshold  SNR$_{spec}$ for a detection.
\begin{equation}
\text{SNR}_{det}=15\sqrt{T}\times\sqrt{\frac{C_{v}(\lambda,D_{tel})t_p}{2C_i(d_{pl}/D_*,\alpha, \lambda)}}D_{tel}\times a\bigg(\frac{r_{pl}^2}{d_{pl}^2}\bigg)\Phi(\alpha)\times\sqrt{\frac{\lambda L_{*,\lambda}}{\text{h}cD_*^2}}
\label{SNRdet}
\end{equation}
The first term is simply the exposure time factor, the second term is determined by the telescope and instrumentation set-up with the third by the planet parameters, and the last term the host star and wavelength band in which we look. Determining $ C_i $ and $ C_v $ accurately for ground-based telescopes is difficult with its dependence on simulating adaptive optics and are functions of the telescope aperture size $D_{tel}$, wavelength $\lambda$ as well as C$_i$ depending on the orbital parameters and system distance $D_*$. The  $ t_p $ term is the telescope throughput. $ L_{*,\lambda} $ the luminosity at the observed wavelength, will be roughly the same for a given spectral type. From this equation we estimate the maximum distance $ D_* $, planet-star separation $d_{pl} $, and planet radius $r_{pl} $ for a given spectral type and thus determine the explorable parameter space of the technique on the ELT . The following analysis assumes a bandwidth of 0.756$\mu$m to 0.773$\mu$m and that the structure/lines in the stellar spectra are sufficient for the above SNR$ _{det} $ results for planet detection to be used.  In reality the structure in the stellar spectra varies significantly with stellar spectral type compared to the M-dwarf case that we extrapolate from here. This would affect the planet detections which make use of the reflected stellar lines. Particularly A and F stars have fewer lines and thus for planet detection would require longer exposure times or to consider a different spectral range given they have fewer features for cross correlation. \\\\Using spectral type magnitude data from \cite{2013ApJS..208....9P} to estimate $ L_{*,\lambda} $ for the O$_2$ A-band spectral range, we run calculations for the ELT . The 0.67" zenith seeing FWHM contrast curve was used to determine the types of planets that could potentially be characterised and/or detected in an exposure time of 50 hours, with a SNR$ _{det}$ of 0.5 and with observations around the O$_2$ A-band wavelength. This SNR$ _{det}$ is sufficient for both planet detection and O$_2$ detection so the results of the calculations are applicable to both.  Clearly, the characterisation of O$_2$ is unlikely to be relevant in the characterisation of Neptunes or Jupiters given the likely temperature regimes and atmospheric compositions of such planets. For these cases in the context of characterisation, the results should be taken as qualitative illustrations of the potential of the technique for characterising such planets in reflected light for those species with strong spectroscopic signatures that are expected to be present. The results are presented in figure 9 below with the colour scale ($ \log_{10}(r_{pl}/R_E) $) indicating $ r_{pl} $ -- the minimum planet radius that can be discovered/characterised for a given orbital separation and distance to the system. Contours for Earth (red), Neptune ($ R_N $) and Jupiter ($ R_J $) sized planets are shown along with the habitable zones (green dashed) of the different spectral types. The habitable zones were calculated as the circumstellar region with a blackbody equilibrium temperature between 273K and 373K.\\\\ The results show that radius contours of the exoplanets that can be discovered have a structure reflecting that of the PSF, with the bumps corresponding to the light and dark rings seen in the PSF used. The bands the bumps trace out represent lines of constant angular separation. The plots show the contours move towards the top left (corresponding to closer systems with less planet-star separation) as the luminosity of the stars decrease towards cooler spectral types. It is important to note that the habitable zone radii contract at a different rate to the detectable radii contours with decreasing stellar luminosity; we find close-by Super-Earth size planets could be detected/characterised for A, F and some M stars, but not for the G and K stars in between. However even the closest A and F stars (Sirius A and Procyon A) are distant enough to require longer exposure times. Exposure times above 50 hours would have a similar effect to considering more luminous spectral types, thus moving the contours out towards the bottom right allowing for smaller planets to be detected/characterised further away with larger separations.\\\\All the plots suggest that the CC method could open up areas of parameter space for which exoplanet systems are believed to exist but currently remain unobservable. Examples include exoplanets similar to Jupiter around nearby G stars (such as $ \alpha$ Cen A), as well as super-Earths and Neptune sized planets that are currently too old, cold and close to their stars for direct imaging in thermal emission, but too far out for transit or radial velocity detection methods. In these conclusions we note the prior assumption that for planet detections this analysis relies on stellar spectra have enough lines/structure for CC such that the SNR$_{det}$ results from above can be used.  We also note that the calculations show the method is heavily restricted to the most nearby systems which is a simply the result of needing bright systems in order to collect enough photons even with the large collecting areas of the next generation ELTs. Current direct imaging attempts also require multiple epoch observations in distinguishing between potential planets and background sources to identify orbital motion or common proper motion. This method may provide another way to resolve this confusion by giving spectra of planet candidates and information on potential orbital motion.  

\begin{figure*}
	\centering
	\includegraphics[width=\linewidth]{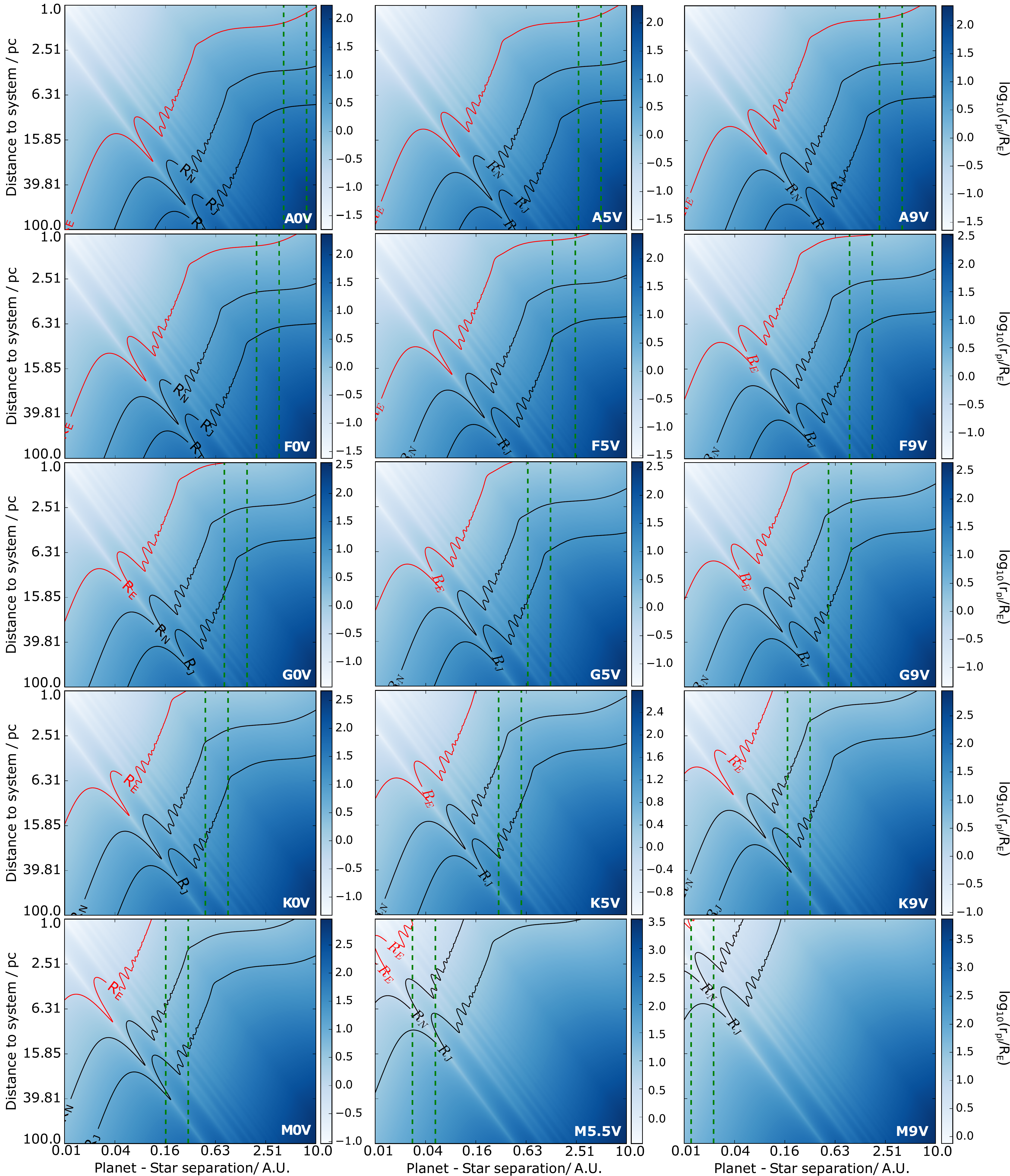}
\label{fig: spectypes}
\caption{The minimum detectable planet radius as a function of distance and separation for stars of various spectral types assuming an exposure time of 50 hours, SNR$_{det}$=0.5 and zenith seeing FWHM = 0.67. The radius also corresponds to the minimum characterisable planet radius for O$_2$ in those planets with the appropriate atmospheric temperatures and compositions (restricted to Earths/super-Earths), though the results are suggestive that other larger planets could be characterised for other spectroscopically strong molecules using this technique. Contours are shown for Earth (red, $R_E$), Neptune ($R_N$) and Jupiter ($R_J$) radii. The habitable zones are shown by the green dashed lines. Note as expected, given the exposure time, the Proxima b-like case (M5.5V) is on the threshold of detectability.}
\end{figure*}

\section{Conclusions}

In conclusion, we determine it is likely the ELT could detect O$ _2$ in the atmosphere of Proxima b by CC of HRS in a reasonable timescale of around 55 hours for R=150,000. This is reliant on the AO systems delivering a PSF equivalent or superior to the 0.67" FWHM curve seen in figure \ref{fig: ELT Ci vs r}  and assumes an orbital phase of 0.5. For instrument contrasts of $ \sim 10^{-2} $ and a $C_v$ value consistent with a Strehl ratio of $ \sim0.3 $ then a method to subtract 99.99\% of the stellar light is required. For higher contrasts, hopefully attainable using HCI systems with XAO, these requirements are relaxed with an instrument contrast of $\sim 10^{-5} $ needing a subtraction method accurate to 99\% and $C_v$ values implying Strehl ratios of $ \sim 10^{-1} $ to $ 10^{-2} $. This work thus gives a good guide to the required performance for the HCI and XAO systems to be coupled to high resolution spectrographs for CC of spectra by the ELT .\\\\ In the extended analysis we show that the  planet spectrum signal-to-noise, SNR$_{spec}$, required for both planet and atmospheric O$ _2 $ detections are significantly reduced, hence exposure times are also dramatically reduced. Detections in the CCF can be made with  SNR$_{CC}$=3 with  the planet spectrum having an  SNR$_{spec}$=0.4 to 1.2 from R=150,000 to R=20,000 for accurate subtractions (99.99\%) of the stellar light.  As noted in \cite{2015A&A...576A..59S} studies of Hot Jupiters using HRS alone suggest removal of stellar light with this accuracy ($\lesssim 10^{-4}$) is possible, however it remains to be proven using HRS in combination with HCI where accurate subtraction of stellar speckles is more difficult. A recent study using medium resolution integral field spectroscopy looks to have achieved accurate stellar light removal of between $\sim 10^{-2}$ to $10^{-3}$ \citep{hoeij2018}. \\\\Finally, in exploring other systems, further analysis of the CC technique suggests the ELT could expand the observable parameter space in terms of exoplanet sizes and separations for nearby systems. Scaling the stellar luminosity from the Proxima b-like case to investigate other spectral types suggests the CC technique could potentially characterise O$ _2 $ for exoplanets around the closest A and F stars, though a full treatment using appropriate stellar spectra is needed for confirmation. We note that although our analysis was limited to constant RV shifts for the parameters of Proxima b and limited O$_2$ A-band wavelength range, the power of the technique demonstrated here will likely hold for many other scenarios of different changing RVs and other spectral features. Next generation ELTs should make use of the technique and thus ensure their high resolution spectrographs have the potential to be coupled with high contrast imaging and integral field instruments. Such methods could also be implemented on current VLTs \citep{lovis2017}.

\section*{Acknowledgements}
We are grateful to the referee for their help in improving the manuscript and thank Matteo Brogi and Jayne Birkby for useful comments on the original thesis. GAH acknowledges the support of an STFC PhD Studentship in writing the manuscript.
% \section*{Acknowledgements}

% The Acknowledgements section is not numbered. Here you can thank helpful
% colleagues, acknowledge funding agencies, telescopes and facilities used etc.
% Try to keep it short.

%%%%%%%%%%%%%%%%%%%%%%%%%%%%%%%%%%%%%%%%%%%%%%%%%%

%%%%%%%%%%%%%%%%%%%% REFERENCES %%%%%%%%%%%%%%%%%%

% The best way to enter references is to use BibTeX:
%\bibliographystyle{mnras}
%\bibliography{hiresproxima} % if your bibtex file is called example.bib

%%%%%%%%%%%%%%%%%%%%%%%%%%%%%%%%%%%%%%%%%%%%%%%%%%

%%%%%%%%%%%%%%%%% APPENDICES %%%%%%%%%%%%%%%%%%%%%

% \appendix

% \section{Some extra material}

% If you want to present additional material which would interrupt the flow of the main paper,
% it can be placed in an Appendix which appears after the list of references.

% %%%%%%%%%%%%%%%%%%%%%%%%%%%%%%%%%%%%%%%%%%%%%%%%%%

% Don't change these lines
\bsp	% typesetting comment
\label{lastpage}
\end{document}